\documentclass[twocolumn]{aastex63}

\usepackage{graphicx}

\newcommand{\Ha}{$\rm H\alpha$ }

\usepackage{bm}
\usepackage{color}

\submitjournal{ApJL}

\shorttitle{Noisy line intensity maps}
\shortauthors{Moriwaki et al.}

\begin{document}

\shorttitle{De-noising LIM with cGAN}
\title{Deep Learning for Line Intensity Mapping Observations: 
Information Extraction from Noisy Maps}

\correspondingauthor{Kana Moriwaki}
\email{kana.moriwaki@phys.s.u-tokyo.ac.jp}

\author{Kana Moriwaki}
\affiliation{Department of Physics, The University of Tokyo, 7-3-1 Hongo, Bunkyo, Tokyo 113-0033, Japan}

\author{Masato Shirasaki}
\affiliation{National Astronomical Observatory of Japan (NAOJ), Mitaka, Tokyo 181-8588, Japan}
\affiliation{The Institute of Statistical Mathematics, Tachikawa, Tokyo 190-8562, Japan}

\author{Naoki Yoshida}
\affiliation{Department of Physics, The University of Tokyo, 7-3-1 Hongo, Bunkyo, Tokyo 113-0033, Japan}
\affiliation{Kavli Institute for the Physics and Mathematics of the Universe (WPI), UT Institutes for Advanced Study, The University of Tokyo, 5-1-5 Kashiwanoha, Kashiwa, Chiba 277-8583, Japan}
\affiliation{Institute for Physics of Intelligence, School of Science, The University of Tokyo, 7-3-1 Hongo, Bunkyo, Tokyo 113-0033, Japan}

\begin{abstract}

Line intensity mapping (LIM) is a promising observational method to probe large-scale fluctuations of line emission from distant galaxies. Data from wide-field LIM observations allow us to study the large-scale structure of the universe as well as galaxy populations and their evolution.
A serious problem with LIM is contamination by foreground/background sources and various noise contributions. 
We develop conditional generative adversarial networks (cGANs) that extract designated signals
and information from noisy maps.
We train the cGANs using 30,000 mock observation maps with assuming a Gaussian noise matched to the expected noise level of NASA's SPHEREx mission.
The trained cGANs successfully reconstruct \Ha emission from galaxies at a target redshift from observed, noisy intensity maps. Intensity peaks with heights greater than
$3.5~\sigma_{\rm noise}$ are located with 60 \% precision.
The one-point probability distribution and the power spectrum
are accurately recovered even in the noise-dominated regime. However, the overall reconstruction 
performance depends on the pixel size and on the survey 
volume assumed for the training data.
It is necessary to generate training mock data with a sufficiently large volume in order to reconstruct the intensity power spectrum at large angular scales. 
The suitably trained cGANs perform robustly against variations of the galaxy line emission model.
Our deep-learning approach can be readily applied to observational data with line confusion and with noise.
\end{abstract}

\keywords{%
high-redshift galaxies ---
large-scale structure of the universe ---
observational cosmology
}

\section{Introduction} \label{sec:intro}
The large-scale structure of the universe contains rich information on galaxy formation and on the nature of dark matter and dark energy.
Line intensity mapping (LIM) is an emerging observational technique that measures the fluctuations of line emission from galaxies and intergalactic medium. 
With typically low angular and spectral resolutions, LIM can survey an extremely large volume.
Future LIM observations are aimed at detecting emission lines at various wavelengths:  H{\sc i} 21cm line \citep[e.g., SKA;][]{Koopmans15}, FIR/submillimeter lines such as [C{\sc ii}] and CO \citep[e.g., TIME;][]{Crites14}, 
and ultraviolet/optical lines such as Ly$\rm \alpha$ and \Ha \citep[e.g., SPHEREx;][]{Dore14}.

While LIM has the advantage of being able to detect all contributions including emission from faint, dwarf galaxies,
there is a serious contamination problem, the so-called line confusion.
Because individual line sources are not resolved in LIM observations, 
foreground/background contamination cannot be easily removed.
So far, only a few practical methods have been proposed to extract designated signals.
Statistics-based approaches include cross-correlation analysis with galaxies/emission 
sources from the same redshift \citep[e.g.,][]{Visbal10},
and one that utilizes the anisotropic power spectrum shape \citep[e.g.,][]{Cheng16}.
\citet{Cheng20} devise a method based on sparsity modeling that successfully reconstructs the positions and the line luminosity functions of point sources from multifrequency data.

Earlier in \citet{Moriwaki20}, we have proposed a deep-learning approach to solve the line confusion problem. We use conditional generative adversarial networks (cGANs), which are known to apply to a broad range of image-to-image translation problems. Our cGANs learn the clustering features of multiple emission sources
and are trained to separate signals from different redshifts.
It is shown that deep learning offers a promising analysis method of data from LIM observations. However, in practice, various noise sources can cause a serious problem.
Faint emission-line signals from distant galaxies are
likely overwhelmed by noise even with the typical level of next-generation observations.

In this Letter, we 
propose to use cGANs to effectively de-noise 
line intensity maps. 
We show that suitably trained cGANs successfully reconstruct the emission line signals on a map,
and recovers basic statistics of the intensity distribution. 
All such information extracted from noisy maps
can be used for studies on cosmology and galaxy population evolution.

\section{Methods} \label{sec:methods}

We consider \Ha emission from galaxies at $z = 1.3$ and observed at 1.5 $\rm \mu m$. 
The \Ha emission is one of the major target lines 
of future satellite missions such as SPHEREx \citep{Dore14} and CDIM \citep{Cooray19}.
We develop cGANs that extract \Ha signals from noisy observational data. 
We first describe how we generate mock intensity maps that are used for training and test.
We then explain the basic architecture of our cGANs.
Further technical details can be found in \citet{Moriwaki20}. 

\subsection{Training and test data}

We prepare a large set of training and test data.
We use a fast halo population code PINOCCHIO \citep{Monaco13} that populates
a cosmological volume with dark matter 
halos in a consistent manner with the underlying 
linear density field.
We generate 300 (1000) independent halo catalogs with a cubic box of $280h^{-1}$ Mpc on a side for training (test)%
\footnote{In the rest of this Letter, we adopt $\rm \Lambda$CDM cosmology with $\Omega_m = 0.32$, $\Omega_\Lambda = 0.68$, and $h = 0.67$ \citep{Planck18}.}. 
The smallest halo mass is $3\times 10^{10}~\rm M_\odot$.
We then assign \Ha luminosities to the individual halos to obtain a three-dimensional emissivity field. 
The halo mass-to-luminosity relation is derived using the result of a hydrodynamics simulation Illustris-TNG \citep{Nelson19}.
We assume that the line luminosity is given by a function of the star-formation rate of the simulated galaxy as
\begin{eqnarray}
	L_{\rm line} = 10^{-A_{\rm line}/2.5} C_{\rm line} \, \dot{M}_*,
\end{eqnarray}
where we adopt $A_{\rm line} = 1.0$ mag,
and $C_{\rm line}$ is a coefficient table computed using the photoionization simulation code Cloudy \citep{Ferland17}.

We work with two-dimensional images (intensity maps) 
in order to make the best use of
modern image translation methods,
although, in principle, it is possible to construct neural networks that read and generate three-dimensional data \citep[e.g.,][]{Zhang19}.
We generate two-dimensional \Ha intensity maps by projecting the three-dimensional emissivity fields along one direction. 

For each realization of the training (test) data, 100 maps (1 map) with an area of $(0.85 ~\rm deg)^2$ are 
generated by projecting random portions of an emissivity field.
A total of 30,000 training data and 1000 test data are generated in this manner.
The intensity maps are pixelized with the angular and spectral resolution of SPHEREx listed in Table \ref{tab1}.

Finally, realistic mock observation maps are generated 
by adding Gaussian noise
to the \Ha intensity map.
We adopt the noise level of "SPHEREx deep" whose $5~\sigma_{\rm n}$ sensitivity per pixel at $\lambda = 1.5\rm \mu$m
is 22 mag, corresponding to $\sigma_{\rm n} = 2.6\times 10^{-6}~\rm erg/s/cm^2/sr$.
The maps are normalized by $1.0\times 10^{-4}~\rm erg/s/cm^2/sr$ before input to the networks.

\begin{table}
\caption{Observational parameters of SPHEREx deep \citep{Dore14}.}
\centering
\begin{tabular}{lc}
    \hline
    Field of view & $200~\rm deg^2$\\
    Angular resolution $l$  & $6".2\times 6".2$\\
    Spectral resolution$\rm ^a$ $R$ & 41.5\\
    Sensitivity$\rm ^a$ $\sigma_{\rm n}$ $[\rm erg/s/cm^2/sr]$ & $2.6\times 10^{-6}$\\ \hline
\end{tabular}
\tablecomments{ $\rm ^a$ Values at $\lambda = 1.5\mu m$.}
\label{tab1}    
\end{table}

\begin{figure*}[ht!]
 \begin{center}
  \includegraphics[clip,width=18cm]{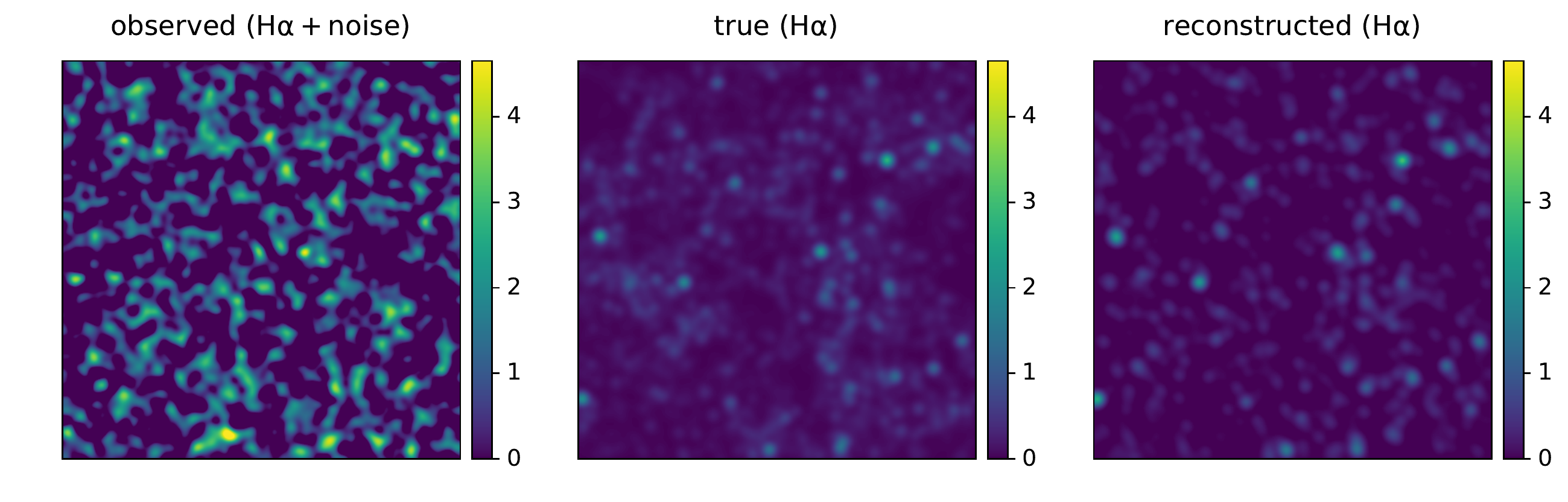}
 \end{center}
   \caption{From left to right, we show the observed map (\Ha + noise), the true \Ha map, and the reconstructed 
   \Ha map. Each map has 0.85 deg on a side. The maps are smoothed with a Gaussian beam with $\sigma = 0.6$ arcmin for 
   better visibility. 
   The intensities are in units of $\rm 10^{-7}~erg {\rm s}^{-1} {\rm cm}^{-2} {\rm sr}^{-1}$.
   Note that the observed map is noise-dominated. See the color-bars on the right.}
   \label{fig:pic}
\end{figure*}

\subsection{Network architecture}

We develop cGANs using the publicly available pix2pix code \citep{Isola16}.
The cGANs consist of two adversarial convolutional networks: a generator and a discriminator.
The generator, consisting of 8 convolutional and 8 de-convolutional layers,
outputs
a map $G$ ("reconstructed map") from an observed map $x$.
The discriminator, consisting of 4 convolutional layers, 
returns a value $D$ for the input of $(x, y)$ or $(x, G[x])$ with $y$ denoting
the \Ha map.
The value $D$ indicates the probability that the input is not $(x,G[x])$ but $(x,y)$.
During the training, the two networks are updated repeatedly
in an adversarial way;
the generator is updated so that it deceives the discriminator (i.e., $D(G[x])$ should get closer to 1), while the discriminator is updated so that it gets better accuracy (i.e., $D(x,y)$ and $D(x,G[x])$ get closer to 1 and 0, respectively).

Specifically,
the parameters in the generator (discriminator) are updated 
to decrease (increase) the loss function
\begin{equation}
    \mathcal{L} = \mathcal{L}_{\rm cGAN}(G, D) + \lambda \mathcal{L}_{\rm L1}(G),
\end{equation}
where
\begin{eqnarray}
    \mathcal{L}_{\rm cGAN}(G, D) &=& \log D(x, y) + \log [ 1-D(x, G[x])],\\
    \mathcal{L}_{\rm L1} &=& \frac{1}{N_{\rm pix}} \sum |y - G(x)|.
\end{eqnarray}
Note that we include an additional term $\mathcal{L}_{\rm L1}$ that is known to ensure better performance 
by imposing the condition that the values of the corresponding pixels in the true and reconstructed maps should be close
\citep{Isola16}.
In each round of training, the loss function is computed with a mini-batch.
After some experiments, we set $\lambda = 1000$ and batch size 4.
The networks are trained for 8 epochs. We adopt these parameter values
throughout the present study.

\section{Results} \label{sec:results}

Figure \ref{fig:pic} shows the reconstruction performance of our cGANs.
Our networks reduce the noise and 
successfully extract the true \Ha signals. 
It is remarkable that both the source positions and the intensities 
are reproduced well, even though the observed map is noise dominated.
A simpler approach in such a noise-dominated case would be to select 
only high signal sources in an observed map. 
However, we find that, if we select pixels with signals greater 
than $3.5 \sigma_{\rm n}$ from 
the observed maps, only 20 \% of them 
are true sources 
(see also Figure \ref{fig:pdf}).
With our networks, about 60 \% of the reconstructed pixels with intensities greater than $3.5 \sigma_{\rm n}$ are real sources. Hence our method significantly outperforms 
the simple
signal selection based on the local intensity.

\subsection{Probability distribution function}

The probability distribution function (PDF) of 
line intensity is an excellent statistic
that can constrain galaxy populations and 
their physical properties
\citep[e.g.,][]{Breysse17}. 
We test whether our networks also recover the PDF accurately.

We first note that, in general, 
a single set of networks do {\it not} 
reproduce pixel statistics of images/maps robustly.
We thus resort to training multiple networks 
and take the mean of the statistics reconstructed
by the {\it ensemble} of networks.
This technique, called "bagging," is known to reduce generalization errors \citep{Goodfellow16},
and has been applied to, for instance, 
de-noising weak lensing convergence maps
\citep{Shirasaki19}.
In practice, we average the PDFs reconstructed by 5 networks that are trained with different datasets.

\begin{figure}[ht!]
\begin{center}
    \includegraphics[clip,width=8cm]{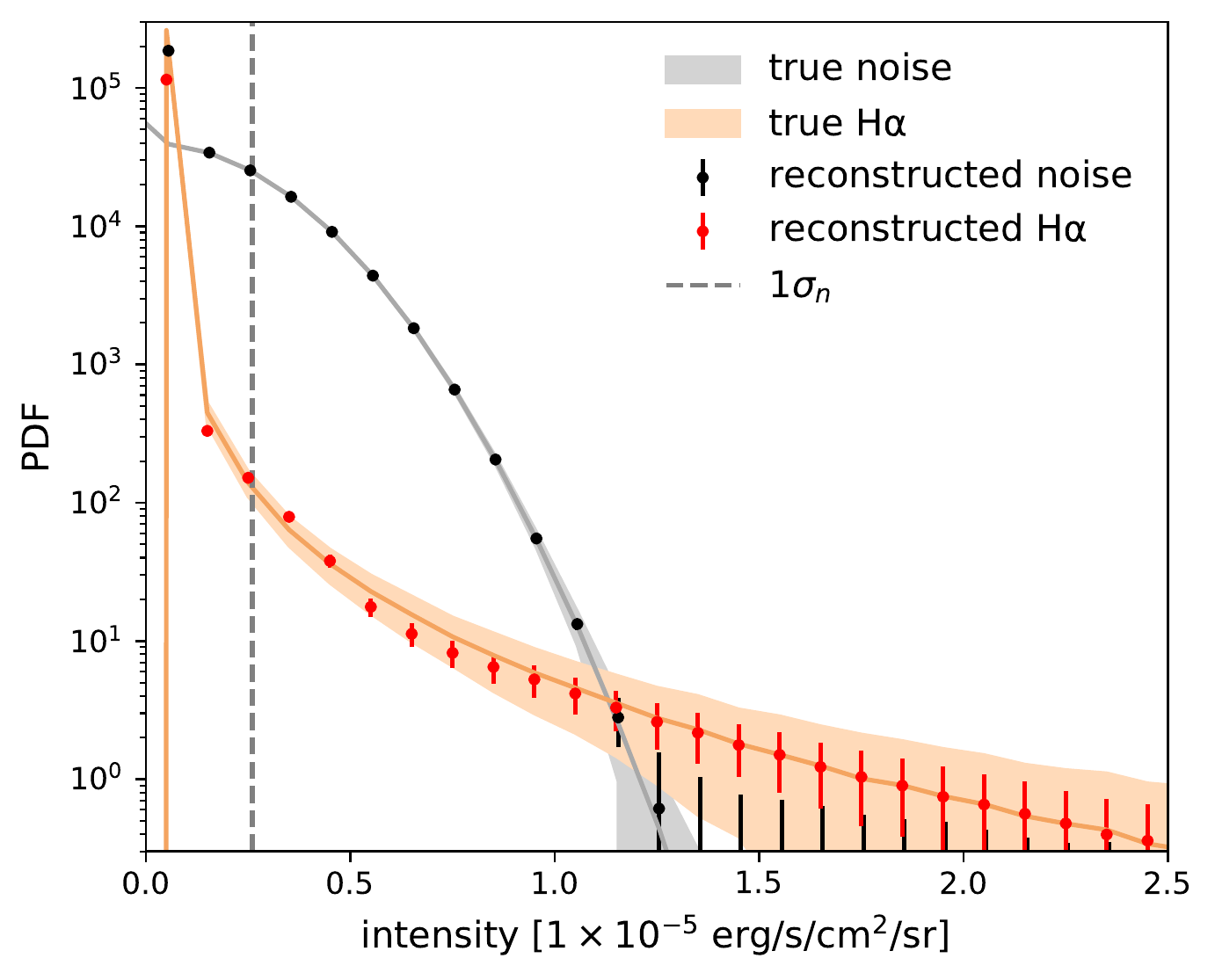}
\end{center}
\caption{The probability distribution function of 
\Ha (red) and noise (black) maps.
The shaded regions and the error bars indicate 1$\sigma$ variances 
of the true and the reconstructed maps evaluated using 1000 test data.
The PDF of a reconstructed map is computed by taking an average of PDFs of 5 reconstructed maps with 5 different networks.
}
\label{fig:pdf}
\end{figure}

Figure \ref{fig:pdf} compares the PDFs of true and reconstructed maps. 
The vertical dashed line indicates the $1\sigma$ noise level.
Our cGANs are able to reconstruct the PDF of the \Ha intensity above 1$\sigma$.
Note also that the scatter of the averaged PDF lies within the intrinsic scatter of the true \Ha maps,
i.e., within the so-called cosmic variance.
Apparently, the networks tend to reconstruct the PDFs 
close to the average. This is simply caused by the bagging
procedure. We have checked and confirmed that 
the variance of reconstructed PDFs 
by a {\it single} network
is as large as the intrinsic one.

\begin{figure}[ht!]
\begin{center}
    \includegraphics[clip,width=8cm]{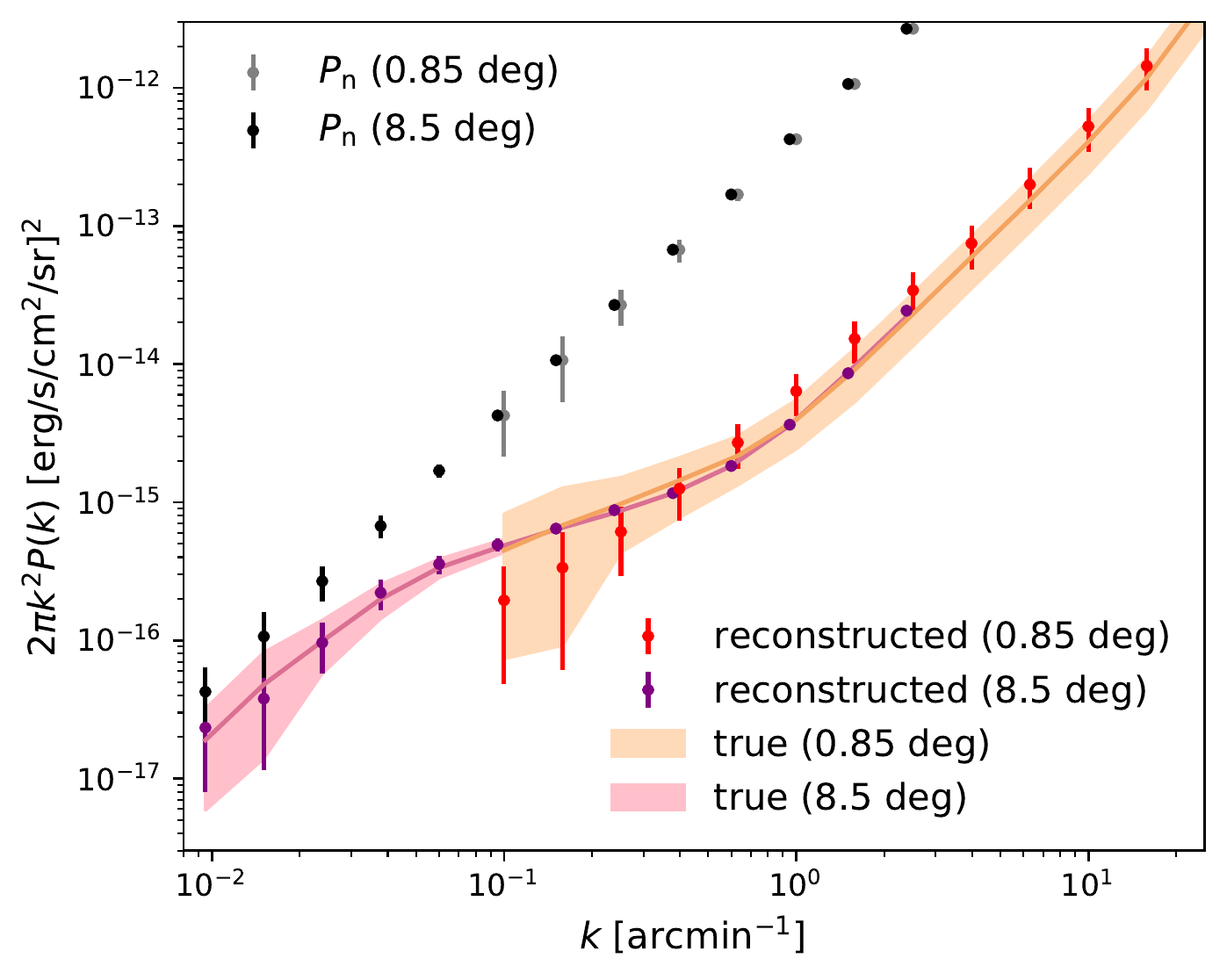}
\end{center}
\caption{The power spectra of \Ha maps.
We show the result of 0.85 deg map (red) and 8.5 deg map (purple).
The shaded regions and error bars correspond to 1$\sigma$ variances 
of the true and reconstructed maps for 1000 test data.
The reconstructed power spectrum is computed by taking an average of those of 
5 reconstructed maps by 5 different networks.
The gray (0.85 deg) and black (8.5 deg) errorbars show the noise power spectra $P_n$ and their uncertainties $P_n/\sqrt{N_k}$.
}
\label{fig:ps}
\end{figure}

\begin{figure*}[ht!]
 \begin{center}
  \includegraphics[clip,width=18cm]{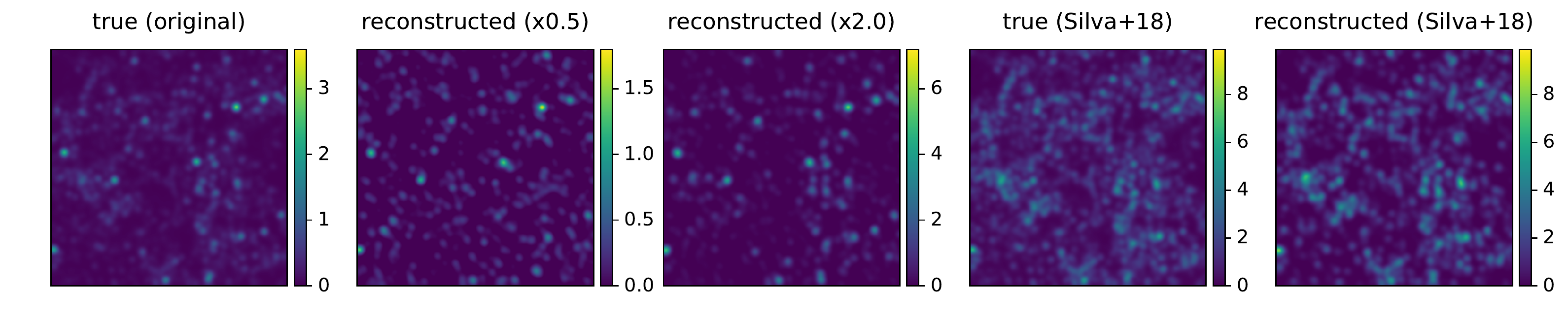}
 \end{center}
   \caption{
   The leftmost panel is the one of the true \Ha maps of original test dataset.
   The second left and middle panel show the reconstructed \Ha maps for test data in which the \Ha intensities are scaled with $c_{\rm H\alpha} = 0.5$, and 2.0, respectively.
   The color bars are scaled so that their corresponding "true" images look the same as the leftmost image.
   The two panels on the right are the true and reconstructed maps 
   when we adopt the \Ha model of \citet{Silva18}.
   Each map has 0.85 deg on a side. 
   All the maps are smoothed with a Gaussian beam with $\sigma = 0.6$ arcmin for better visibility.
   The intensities are in units of $\rm 10^{-7}~erg {\rm s}^{-1} {\rm cm}^{-2} {\rm sr}^{-1}$.
   }
   \label{fig:pic-test}
\end{figure*}


\subsection{Power spectrum}

We further examine the ability of our cGANs to reconstruct the intensity power spectrum.
To this end, we again adopt the bagging of 5 networks
that are trained with different datasets.
The red points with error bars and the shaded regions in Figure \ref{fig:ps} 
show the power spectra of the reconstructed and true \Ha maps, respectively. 
We also show the noise power spectrum, $P_n (k)$, and its variance, $P_n (k)/\sqrt{N_k}$,
where $N_k$ is the number of modes in 
$k-\Delta k/2 < |\bm{k}| \leq k + \Delta k/2$.
We adopt $\Delta \log k = \Delta k /k = 0.2$.

We notice that
the reconstructed power spectra on large scales ($k \lesssim 0.5$) are systematically underestimated.
This might be owing to the finite box size of the training data.
To examine this, we train the cGANs with mock intensity maps with
a larger area. For this test, we generate halo catalogs in a cubic volume of 
$700h^{-1}$ Mpc on a side (see Section 2.1).
Then the smallest halos populated is 
degraded to 
$3\times 10^{11}~\rm M_\odot$,
but we have confirmed that the mean \Ha intensity (or the total luminosity density) is not significantly different from that with our default 
box size of $280h^{-1}$ Mpc. 
We set the side length of the pixel $l' = 2.0 ~\rm arcmin$ and the spectral resolution $R' = 41.5$.
Each map has a ten times larger area of $(8.5~\rm deg)^2$.
The noise level scales with the angular and spectral resolution as
\begin{eqnarray}
    \sigma_{\rm n}' = \sigma_{\rm n} \sqrt{\Big(\frac{l}{l'}\Big)^2 \frac{R'}{R}},
\end{eqnarray}
where $l$, $R$, and $\sigma_{\rm n}$ are the original angular and spectral resolution and the noise level of SPHEREx (Table \ref{tab1}).
The resulting noise level of the wide maps is 
$\sigma_{\rm n}' = 1.3 \times 10^{-7}~\rm erg/s/cm^2/sr$.
We adopt the same hyperparameters in the cGANs as in our default case except we set $\lambda = 200$ and 
the normalization factor $1.0\times 10^{-6}~\rm erg/s/cm^2/sr$ 
for the low-resolution maps.
The purple dots with error bars in Figure \ref{fig:ps} show 
the power spectrum of the reconstructed wide maps. The light-pink shading 
indicates the 1$\sigma$ dispersion of the true \Ha
power spectra.
Clearly, the large-scale (low-$k$) power spectrum is reconstructed 
more accurately compared to our default case.
We note that the cGANs trained with the wider maps do not resolve 
point sources, but the peaks and voids in the reconstructed map
correspond closely to the positions of groups/clusters and void regions.

Ideally, networks trained with intensity maps that have fine pixels and a large box size would 
be able to 
reconstruct both the positions of point sources (galaxies) and their large-scale clustering.
Unfortunately, 
it becomes computationally more expensive 
if we set a larger number of pixels. The computational time for training roughly scales with the number of pixels, and the 
necessary number of training epochs could also increase.
We thus suggest that one should generate training data depending on the purpose.
In order to detect point sources robustly, one needs to train the networks using fine-pixel maps.
If the primary purpose is to reconstruct the large-scale power spectrum, for cosmology studies for instance, 
then one needs to generate maps with a sufficiently large area (volume) but with coarse pixels. 
The reconstructed power spectra shown in Figure \ref{fig:ps} suggest that one should adopt at least 
a several times larger area for training than the actual 
size of observed maps.

\subsection{Line emission models}

\begin{figure*}[ht!]
\begin{center}
    \includegraphics[clip,width=15cm]{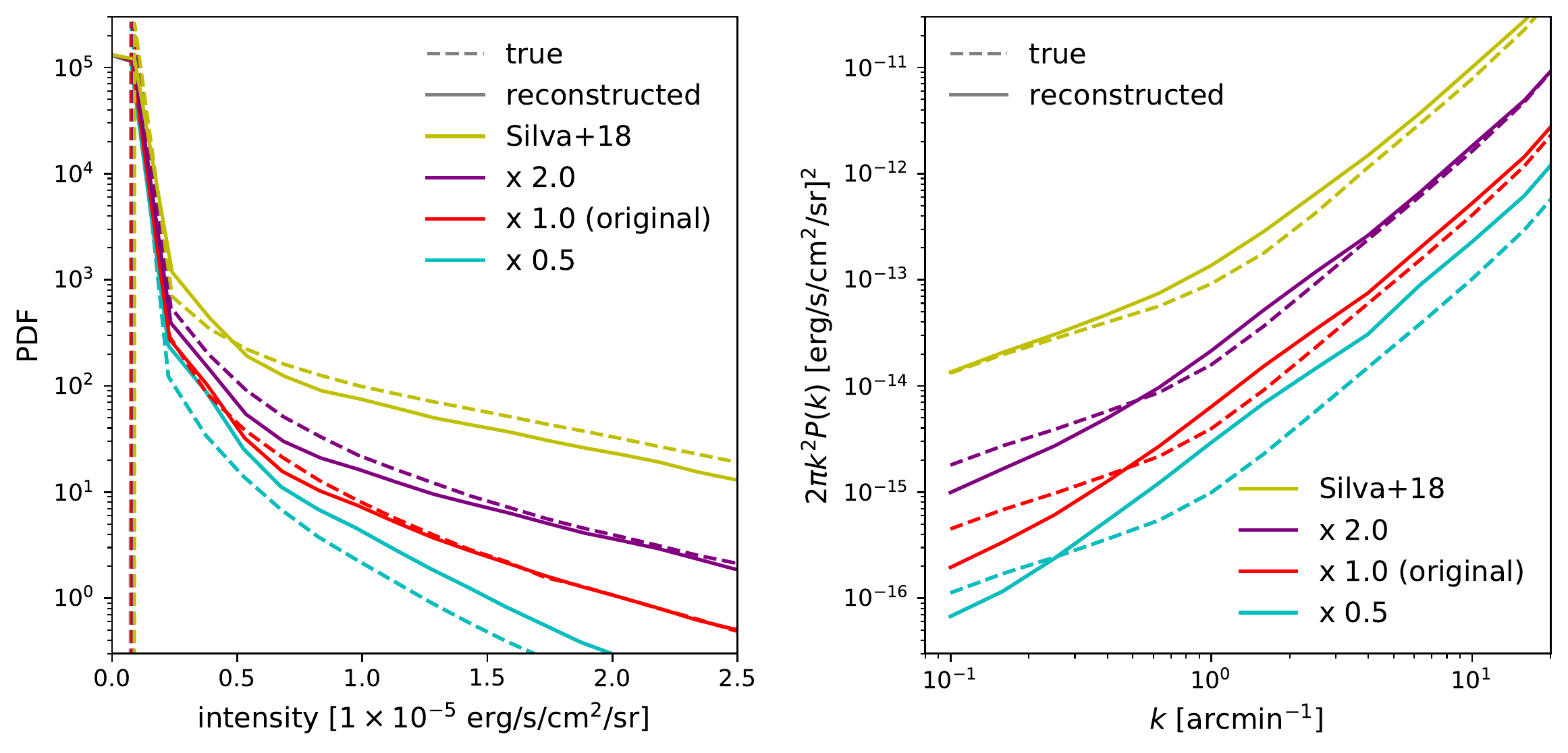}
\end{center}
   \caption{
   The average of the probability distribution function (left) and the power spectrum (right) 
   of the true (dashed) and the reconstructed (solid) \Ha maps over 1000 test dataset.
   The reconstructed statistics are computed by taking means of those of 5 different networks.
   The cyan, red, purple, and yellow lines correspond to the results with $c_{\rm H\alpha} = 0.5, 1$, 2, and \citet{Silva18} model, respectively.
}
\label{fig:pdf-ps-test}
\end{figure*}

The line intensities of high-redshift galaxies are not well constrained observationally, and theoretical models of galaxy formation and evolution remain uncertain. Thus it is important to examine whether our GAN-based method can be applied robustly to data that are generated with different line emission models.
To this end, we generate three additional sets of 1000 test data that have the same noise level but have different \Ha line intensities.
Two of the new test datasets are generated by simply multiplying the original \Ha maps by a constant value, $c_{\rm H\alpha} = 0.5$ and 2, as
\begin{eqnarray}
    I_{\rm obs}(\bm{x}) = c_{\rm H\alpha} I_{\rm H\alpha}(\bm{x}) + I_{\rm noise}(\bm{x}).
\end{eqnarray}
These cases serve as a test of the networks' reconstruction performance against the variation 
in the mean intensity.
We also adopt a completely different \Ha model of emission, \citet{Silva18}, in which 
the SFR-halo mass relation derived by \citet{Guo13}
is used to assign the \Ha luminosities ($\propto$ SFR). The model reproduces the mean \Ha intensity
consistent with observations \citep{Sobral13, Sobral15}.

Figure \ref{fig:pic-test} shows examples of reconstructed images.
In all the cases, the locations and amplitudes of bright peaks are reconstructed well,
even though the mean intensity levels are quite different from the original data (see the color bars).
It appears that the networks practically learn the {\it noise properties} and extract the signal robustly, even if the signal amplitude is different from 
the training data.
It is remarkable that,
in the cases with $c_{\rm H\alpha} = 2$ and with \citet{Silva18} model, the \Ha signals are  reconstructed as accurately as the default 
$c_{\rm H\alpha} = 1$ case.

We also compare the one-point PDFs and the power spectra in Figure \ref{fig:pdf-ps-test}, where
we plot the averages over 1000 test data with $c_{\rm H\alpha} = 0.5$ (cyan line), 1 (red), and 2 (purple), respectively. The yellow lines show those with the \citet{Silva18} model. 
The statistics of each reconstructed image are computed by taking the means of those reconstructed by 5 different networks.
Both the one-point PDFs and the power spectra are reproduced well except the case of $c_{\rm H\alpha} = 0.5$ that has effectively small signals.
We thus conclude that our network is applicable to 
maps with different intensity models as long as we have a good understanding of the observational noise and if the line intensities are not too weak compared to the training data.

\section{Conclusion}

We have developed cGANs that effectively reduce  observational noise in line intensity maps.
We train the cGANs by using a large set of mock observations assuming 
a realistic noise level expected for the SPHEREx mission.
Our cGANs can reconstruct the point-source positions and the PDF of the intensity maps.
The power spectrum is also reconstructed remarkably well, but the accuracy depends on the area/volume 
assumed for the training data. 
We have also found that our method is able to reconstruct the signals even if the underlying line intensity model 
is different from the original training data.

If we combine with another set of networks that efficiently separates signals from different redshifts \citep{Moriwaki20}, the cGANs developed in this study can extract the emission-line signal from galaxies 
at an arbitrarily specified redshift from noisy maps. 
Therefore, using data from multi-frequency, wide-field intensity mapping observations, we can reconstruct the
three-dimensional distribution of emission-line galaxies. 
The intensity peaks detected by our cGANs
correspond to bright galaxies and galaxy groups 
with high confidence, which will be promising targets 
for follow-up observations.
Finally, the reconstructed line intensity map essentially traces the distribution of galaxies and hence of underlying matter,
and thus it is well suited for cross-correlation analysis with other tracers.
Accurate reconstruction of the statistics such as the one-point PDF and power spectrum as shown in this Letter will allow us to
perform cosmological parameter inference 
and to study galaxy formation and evolution
using data from future LIM observations.

\acknowledgments

We thank the anonymous referee for helpful comments and suggestions.
We thank Yasuhiro Imoto for helpful comments.
KM is supported by JSPS KAKENHI Grant Number JP19J21379, by Advanced Leading Graduate Course for Photon Science (ALPS) of the University of Tokyo,
and by JSR Fellowship, the University of Tokyo.
MS is supported by JSPS Overseas Research Fellowships. 
NY acknowledges financial support from JST CREST JPMHCR1414
and JST AIP Acceleration Research JP20317829.

\software{Cloudy \citep{Ferland13}, Pinocchio \citep{Monaco13}}

\bibliography{bibtex_library}{}
\bibliographystyle{aasjournal}

\end{document}